\documentclass[journal]{IEEEtran}

\usepackage{xcolor,soul,framed} 

\colorlet{shadecolor}{yellow}
\usepackage[pdftex]{graphicx}
\graphicspath{{../pdf/}{../jpeg/}}
\DeclareGraphicsExtensions{.pdf,.jpeg,.png}

\usepackage[cmex10]{amsmath}
\usepackage{array}
\usepackage{mdwmath}
\usepackage{mdwtab}
\usepackage{eqparbox}
\usepackage{url}

\begin{document}

\title{Blockchain Methods for Trusted Avionics Systems}

\author{
\IEEEauthorblockN{Erik Blasch$^{a}$, Ronghua Xu${^b}$, Yu Chen${^b}$, Genshe Chen$^{c}$, Dan Shen${^c}$}

\IEEEauthorblockA{$^{a}$The U.S. Air Force of Scientific Research (AFOSR), Arlington, VA 22203\\ ${^b}$Dept. of Electrical \& Computer Engineering, Binghamton University, SUNY,  Binghamton, NY 13902 \\ 
${^c}$Intelligent Fusion Tech, Inc, Germantown, MD 20876\\
erik.blasch@us.af.mil, \{rxu22, ychen\}@binghamton.edu, \{gchen, dshen\}@intfusiontech.com}
}

\maketitle

\begin{abstract}
Blockchain is a popular method to ensure security for trusted systems. The benefits include an auditable method to provide decentralized security without a trusted third party, but the drawback is the large computational resources needed to process and store the ever-expanding chain of security blocks.  The promise of blockchain for edge devices (e.g., internet of things) poses a variety of challenges and strategies before adoption. In this paper, we explore blockchain methods and examples, with experimental data to determine the merits of the capabilities.  As for an aerospace example, we address a notional example for Automatic dependent surveillance—broadcast (ADS–B) from Flight24 data (https://www.flightradar24.com/) to determine whether blockchain is feasible for avionics systems. The methods are incorporated into the Lightweight Internet of Things (IoT) based Smart Public Safety (LISPS) framework. By decoupling a complex system into independent sub-tasks, the LISPS system possesses high flexibility in the design process and online maintenance. The Blockchain-enabled decentralized avionics services provide a secured data sharing and access control mechanism. The experimental results demonstrate the feasibility of the approach.
\end{abstract}


\IEEEpeerreviewmaketitle

\section{Introduction}
\label{intro}

Air travel has become increasing dense in time (e.g., air traffic management), space (e.g., system wide information management), and frequency (e.g., communications) that places increase burdens on avionics to support public safety and security \cite{blasch2019cyber}. The future requires methods for increased cyber security \cite{do2017game} for avionics systems \cite{blasch2015summary}.

Advances in software, hardware, and communication generate new opportunities for malicious activities (Fig 1). An example is Automatic dependent surveillance—broadcast (ADS–B) which poses challenges for confidentiality, availability, and integrity over security and safety concerns \cite{costin2012ghost, manesh2017analysis, purton2010identification, strohmeier2014security}. Current constructs include the system design as a cyber-physical system (CPS), as well as the devices operating at the edge to include the internet of things (IoT), that require consideration to mitigate cyber vulnerabilities. The continued use of unmanned aerial systems (UAS) increases the need for cyber awareness \cite{cruise2018cyber, kramer2018adaptation, raz2019enabling} for aeronautical informatics \cite{durak2018advances}.

Cyber awareness concerns require new methods for security, such as blockchain, for enhance performance over reliability, resilience, and assurance. Areas of concern include:

\begin{itemize}
    \item \textbf{Networks} – The various fixed and wireless ground and air constructs that enable the delivery of information to and from the aircraft, ground, and space \cite{roy2014aeronautical, shen2007adaptive, shen2014network}. Examples include air traffic management (ATM) with Internet Protocol (IP) addresses.
    
    \item \textbf{Electronics} – The on-board avionics is subject to internal and external performance requirements against size, weight, and power requirements. Examples include the battery power \cite{leuchter2015investigation} and sensors supporting engine control.
    
     \item \textbf{Software} – As modern systems are operating with large data, the control and run-time operations require sophisticated methods for efficiency. Examples include integrated modular avionics (IMA) \cite{gaska2015integrated}.
     
     \item \textbf{Analytics} – Availability, confidence, and processing of systems is determined by the various standards in development and deployment designs that meet performance \cite{alsing19993d, batuwangala2017safety} and effectiveness criteria \cite{yang2014mobile}. Examples include the compliance and mandates for Global Positioning System (GPS), video \cite{blasch2012wide}, or Automatic dependent surveillance—broadcast (ADS–B) \cite{insaurralde2019uncertainty} use. 
     
     \item \textbf{Communication} – A key aspect of cyber is the coordination of the signals that are transferred. For air operations, the wireless signals from the space and air pathways need to operate reliably \cite{xu2018high}. Examples include performance-based navigation and communications, surveillance, and navigation (CNS) signals for coordinating flight.

    \item \textbf{Data} – On the physical networks and communication pathways, the data and protocols should provide information with integrity and consistency \cite{imai2017airplane}. Examples include System Wide Information Management (SWIM) capability for real-time support for collision avoidance.
\end{itemize}

\begin{figure}[t]
    \vspace{-10pt}
    \centering
        \includegraphics[width=0.48\textwidth]{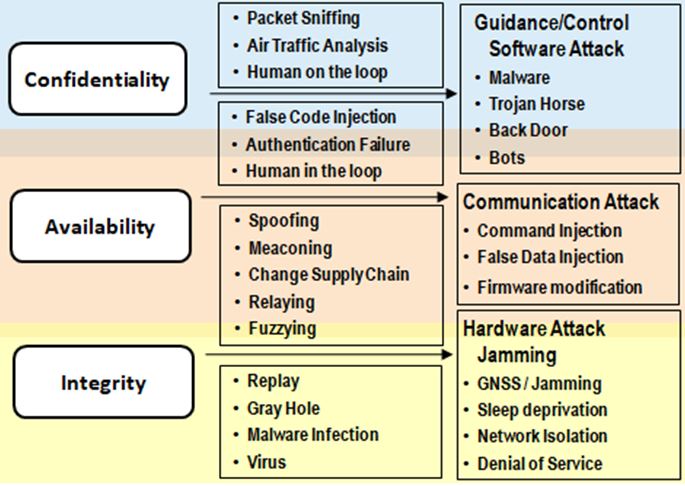}
    \caption{Avionics Vulnerabilities.}
    \label{fig:vulerabilities}
\end{figure}

Figure \ref{fig:vulerabilities} highlights the growing concerns including software, communication, and hardware attacks.The rest of the paper is as follows. Section \ref{background} provides a review of avionics cyber concerns. Section \ref{sys} discusses microservices. Section \ref{data} provides a discussion of the LISPS framework. Section \ref{fuzzy} provides an avionics use case. Section \ref{conc} provides conclusions.

\section{Background Cyber Avionics}
\label{background}

In the last twenty years, the cyber-landscape in ATM has evolved significantly, particularly in conjunction with advances in hardware (e.g., wireless and Internet of things (IoT) devices, networks (e.g., IP-based connectivity from Aeronautical Fixed Telecommunication Networks (AFTN)), software (e.g., SESAR and NextGen technologies, including avionics ontologies \cite{insaurralde2018moral, insaurralde2018uncertainty, insaurralde2018uncertain}, Aeronautical Mobile Airport Communication System (AeroMACS), etc.) \cite{batuwangala2018certification}. The future of aviation includes greater amounts of data exchanged in real-time across an increasing number of stakeholders.

\subsection{Cyber Avionics Access Issues}

Cyber-security was largely inherent in traditional ATM systems thanks to the limited or null interconnectivity between dedicated CNS/ATM subsystems. However, current ATM requires highly interconnected Decision Support Tools (DST) addressing the requirements of both strategic and tactical operational timeframes. Currently, most Air Navigation Service Providers (ANSP) implement the following CNS measures:

\begin{itemize}
    \item \textbf{Access protocols} - Online submission of flight plans and flight plan amendments \cite{homola2019aviation} can only be performed by authenticated and authorized users and is subject to very detailed scrutiny (checks are performed to avoid any intentional/unintentional duplication in flight crew, aircraft, call-sign).
    \item \textbf{Authentication} and user restrictions especially in relation to external entities participating to the collaborative decision making process (airlines, weather offices, handling agents, airport management, etc.) \cite{sabatini2016cyber}.
\end{itemize}

Future techniques of emerging interest include blockchain techniques for access and authentication security \cite{kapitonov2017blockchain, xu2018blendcac}.

\subsection{Cyber Avionics Vulnerabilities}

Encryption and tunnelling (IP-based interconnectivity): increasingly adopted for ground-based telecommunication networks provide measures of development that continual need to be designed to ensure that that methods are not subject to attack. Essential to these methods are the confidence, integrity, and availability (continuity) of the signals (see Fig. 1) being used by avionics systems for effective performance \cite{wei2015effectiveness}. Example threats that can severely affect future systems can be categorized as firmware (software) and network (physical) attacks.  

\noindent \textbf{Firmware Attacks:}

\begin{itemize}
    \item \textbf{Supply Chain Attack}: Gaining access to supplier computers and modifying the firmware (e.g., pre-installing back doors, malicious code, etc.)
    \item \textbf{False Data Injection}: Compromising a computer-controlled sensor by injecting false data in computer-driven data analysis/fusion process.
    \item \textbf{False Code Injection}: Providing addition software routines to alter processing, skip functions, or deny outputs.
    \item \textbf{Firmware Modification}: Interjecting software to alter the processing of low-level device control and the operating system that control data and manipulation functions.
    \item \textbf{Malware Infection}: Inserting software into the system with deliberate harmful intent including viruses, worms, back doors and Trojan horses.
\end{itemize}

\noindent \textbf{Network Attacks:}

\begin{itemize}
    \item \textbf{Jamming}: Transmitting high power signals to impede reception of radio frequency/electro-optical (RF/EO) signals (i.e., degrading accuracy and continuity) \cite{wei2007multi}.
    \item \textbf{Spoofing/Meaconing}: Synthesizing and transmitting a false signal \cite{manesh2019detection} to deceive a target RF/EO sensor’s positioning and/or tracking data; Meaconing refers to capturing legitimate RF/EO signal and rebroadcasting with alterations (e.g., time delay), affecting the RF/EO sensor estimation accuracy, continuity and/or integrity (e.g., GNSS \cite{sabatini2015assessing, yang2016simultaneous}).
    \item \textbf{Denial of Service (DoS)}: DoS attacks can be launched at any layer of an avionics system. In the physical layer, the adversary can launch a jamming attack to interfere with communication on physical channels. In the network layer, the adversary can disrupt the routing protocol and disconnect the network. In the application layer, the adversary can shut down high-level services \cite{wei2014simulation}.
    \item \textbf{Command Injection}: Overtaking control by interjection of access such as Buffer-Overflow Attack \cite{krishna2017review}.
    \item \textbf{False Data Injection}: Manipulating the data such as a data reply attack with available or new information \cite{sabatini2015assessing}.
    \item \textbf{Network Isolation}: Shutting down services to disconnect the proper communication functions such as an ARP (Address Resolution Protocol) Cache Poisoning approach \cite{sabatini2015assessing}.
    \item \textbf{Packet Access}: Sniffing and fuzzifying the packets that contain the data for direct access of to intend for additional information transfer to other systems \cite{sabatini2015assessing}.
\end{itemize}

\subsection{ADS-B}

\begin{figure}[t]
    \centering
        \includegraphics[width=0.3\textwidth]{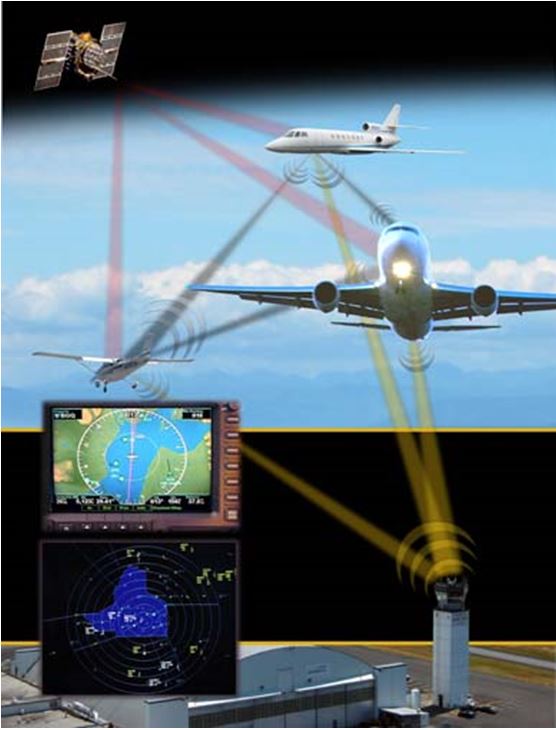}
    \caption{ADS-B (from http://www.faa.gov/nextgen/media/).}
    \label{fig:ads-b}
    \vspace{-10pt}
\end{figure}

The Automatic Dependent Surveillance-Broadcast (ADS-B) for avionics allows aerospace systems to broadcast position, air speed, and identification over a radio message at 1090 MHz (see Fig. \ref{fig:ads-b}). It is lower cost than conventional radar systems, enhances surveillance movement analysis, and increase pilot situational awareness. The system composes 1) ground infrastructure, 2) airborne component, and 3) operating procedures. ADS-B relies on the Global Positing System (GPS) or other systems like the inertial navigation system (INS). Currently, the line-of-sight range is 370 km.

``ADS-B out'' is from the aircraft and provides functionality to determine dynamics of turns and descent/ascent. “ADS-B in” includes radio messages received by air traffic control stations and other aircraft including platform traffic positions, weather updates, and flight information. However, after deployment, there were reports of message spoofing because the data was not encrypted or authenticated \cite{wei2007multi}. Blockchain microservices could be a solution for ADS-B security.


\section{Microservices}
\label{sys}

There is a growing demand for human resources to interpret the live data streams (e.g., from security cameras \cite{blasch2012high, chakravarthy2015adapting, liu2014holistic}). Numerous automated object detection algorithms have been investigated to atomize this process using statistical analysis \cite{fuse2017statistical} or machine learning (ML) \cite{ribeiro2018study} approaches. The ML methods are computationally expensive such as using algorithms that are normally implemented for powerful cloud servers for a surveillance system. For example, Wide Area Motion Imagery (WAMI) streams video from sensors back to the cloud for processing, which puts a heavy burden on the communication network \cite{wu2017container, wu2015pseudo}. To reduce data transmission, it has been suggested to promote operators’ awareness by using context \cite{blasch2014context, snidaro2016context} or providing query languages \cite{aved2015multi}. Further, approaches such as re-configuring the networked cameras \cite{piciarelli2015dynamic}, utilizing event-driven visualization \cite{fan2016heterogeneous}, and mapping conventional real-time images to camera images \cite{liu2014information} improve the efficiency and throughput of the communication networks along with better detection rates.

Decentralized surveillance systems are more suitable in many mission-critical, delay sensitive tasks \cite{nikouei2018eiqis}. Recent developments of the edge hierarchy architecture enables real-time surveillance based on the fog computing paradigm \cite{mukherjee2018survey}. Many on-line and uninterrupted target tracking systems are proposed to meet the requirements of real-time measurement processing and instant decision making deployed at the edge \cite{blasch2000data, dunik2015random, van2018micro, van2015comparative}. Researchers also merged raw data streams from drones on near-site fog computing devices to reduce the amount of data to be outsourced to the cloud center \cite{chen2016dynamic}.

A safety system focusing on object assessment can be constructed following the edge-fog-cloud hierarchy \cite{mouradian2017comprehensive}. The input surveillance video frame is given to an edge unit where low-level processing is performed, such as feature detection and object tracking \cite{blasch1998simultaneous, howard2017mobilenets, nikouei2018smart}. The intermediate-level is in charge of action recognition, behavior understanding, and decision making like abnormal event detection, which is implemented at the fog stratum \cite{nikouei2019kerman, nikouei2018real}. Finally, the high-level is focused on historical pattern analysis, algorithm fine tuning, and global statistical analysis \cite{blasch2010high, blasch2014context, blasch2013revisiting} such as flight routes.

Exploring context services includes that of dynamic data driven applications systems (DDDAS) \cite{blasch2018dddas, blasch2018handbook, fujimoto2018dynamic} gathered from scientific applications: (1) theory (e.g., estimation); (2) methods (e.g., image computing for situation evaluation \cite{zheng2018multispectral, zheng2012qualitative,  zuo2017combining, zuo2017covert}); and (3) design (situation awareness through contextual assessment). Enhance user interaction through a user defined operating picture (UDOP) interface \cite{blasch2000assembling, blasch2006level, blasch2013enhanced, blasch2003situation} for decision making \cite{blasch2011user}.

The next sub-sections introduce the microservice architecture and blockchain technology useful for avionics systems.
 
\subsection{Microservices in IoT}

A Service Oriented Architecture (SOA) is widely adopted in the development of application software in IoT and CPS environments \cite{butzin2016microservices}. The traditional SOA utilizes a monolithic architecture that constitutes different software features in a single interconnected and interdependent application database. Owing to the tightly coupled dependence among functions and components, such a monolithic framework is difficult to adapt to new requirements for IoT-enabled avionics systems, such as scalability, service extensibility, data privacy, and cross-platform interoperability \cite{datta2018next, du2014garp}. As an extension of the traditional SOA, a microservices architecture allows functional units of an application to work independently with a loose coupling by encapsulating a minimal functional software module as a microservice, which can be individually developed and deployed. Each microservice is a process dedicated to certain function of the application. The individual microservices asynchronously communicate with each other through a lightweight mechanism, such as an HTTP RESTful API or a message bus \cite{lu2017secure}. Finally, multiple decentralized individual microservices cooperate with each other to perform the functions of complex systems. The flexibility of microservices enables continuous, efficient, and independent deployment of application functional units.

Thanks to these attractive properties, such as fine granularity and loose coupling, the microservices architecture has been investigated in many smart solutions to enhance the scalability and security of IoT-based applications. Current IoT systems are advancing from “things”-oriented ecosystems to a widely and finely distributed microservices-oriented ecosystems with potential for avionics navigation. An Intelligent Transportation Systems (ITS) that incorporates and combines data approaches using the serverless microservices architecture has been designed and implemented to help the transportation planning for the Bus Rapid Transit (BRT) system \cite{herrera2018smart}. To enable a more scalable and decentralized solution for advanced data stream analysis for large volumes of distributed edge devices, a conceptual design of a robust smart surveillance system was proposed based on microservices architecture and blockchain technology \cite{nikouei2019decentralized}. It aims at offering a scalable, decentralized and fine-grained access control solution for smart surveillance systems.

\subsection{Blockchain-enabled Security}

\begin{figure}[t]
    \centering
        \includegraphics[width=0.48\textwidth]{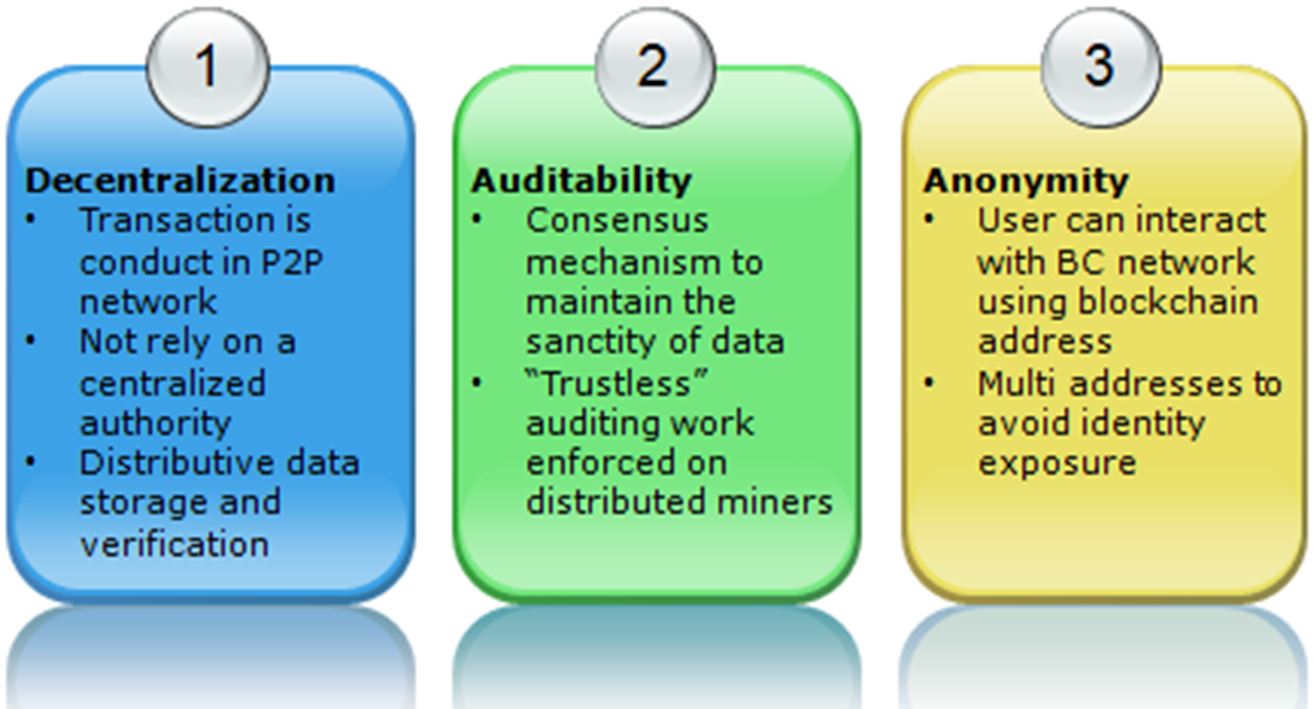}
    \caption{Blockchain Advantages.}
    \label{fig:bc-adv}
    \vspace{-10pt}
\end{figure}

As a fundamental technology of Bitcoin \cite{nakamoto2008bitcoin}, blockchain initially was used for new cryptocurrencies that perform commercial transactions among independent entities without relying on a centralized authority. Essentially, the blockchain is a public ledger based on consensus rules to provide a verifiable, append-only chained data structure of transactions. Due to the decentralized architecture, blockchain allows the data to be stored and updated, which makes blockchain an ideal architecture to ensure distributed transactions among all participants in a trustless environment, like avionics networks (as shown in Fig. \ref{fig:bc-adv}).

The blockchain consensus properties include:
\begin{itemize}
    \item \textbf{Validity (Correctness)}: If a honest node receives a valid common replicate proposed by other nodes, this common replicate should be accepted into the blockchain.
    \item \textbf{Agreement (Consistency)}: All the honest node should updates their local replicates of the blockchain with that block header of confirmed global blockchain.
    \item \textbf{Termination (Liveness)}: Every honest node should either discard or accept new transactions into blockchain. Finally, all transactions originated from the honest nodes will be eventually confirmed.
    \item \textbf{Integrity (Total order)}: All honest nodes should accept the same chronological order of transactions which are correctly appended to hash-chained blockchain. 
\end{itemize}

\begin{figure}[t]
    \centering
        \includegraphics[width=0.46\textwidth]{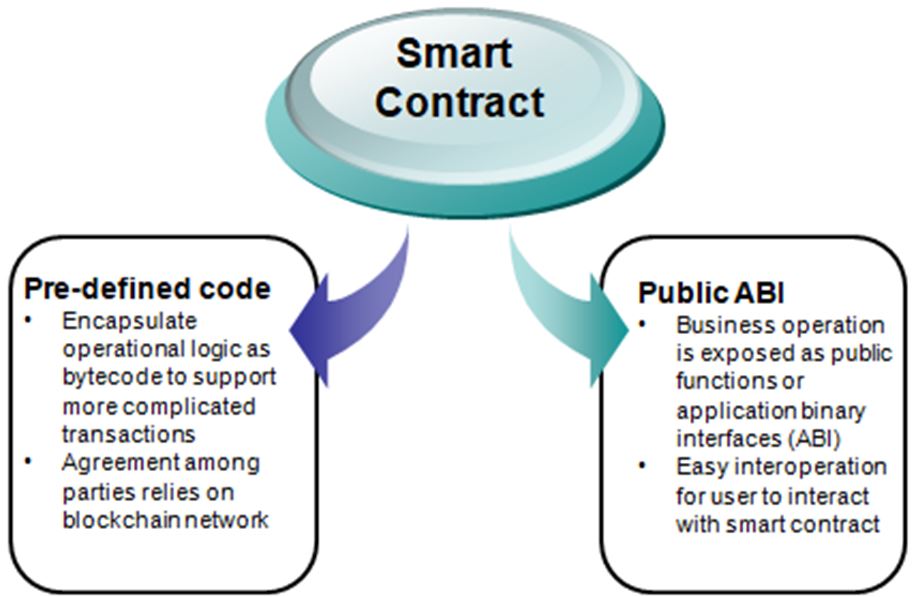}
    \caption{Smart Contract.}
    \label{fig:smart}
    \vspace{-10pt}
\end{figure}

Emerging from the intelligent property, a smart contract allows users to achieve agreements among parties through a blockchain network, Fig. \ref{fig:smart}. By using cryptographic and security mechanisms, a smart contract combines protocols with user interfaces to formalize and secure relationships over networks \cite{szabo1997formalizing}. A smart contract includes a collection of pre-defined instructions and data that have been saved at a specific address of blockchain as a Merkle hash tree, which is a constructed bottom-to-up binary tree data structure. Through exposing public functions or application binary interfaces (ABI), a smart contract interacts with users to offer the predefined business logic or contract agreement. A smart contract enhances trust, efficiency, accuracy, and autonomy.

The blockchain and smart contract enabled security mechanism for applications have been reported recently, for example, smart surveillance system \cite{nagothu2018microservice}, identification authentication \cite{hammi2018bubbles}, access control \cite{jia2016cooperative, xu2018blendcac}, social credit system \cite{xu2018constructing}, biomedical imaging data processing \cite{xu2019decentralized}, and space situation awareness \cite{xu2019exploration, xu2018real}. Thus, blockchain and smart contract together are promising to provide a solution to enable a secured data sharing and access authorization in decentralized avionics systems.

\section{System Architecture}
\label{data}

The proposed microservices enabled blockchain network for secure avionics framework follows the divide-and-conquer principle to functionally decouple the processes for navigation and system security. The computationally expensive processes are divided into multiple sub-tasks. Based on the microservices architecture, the proposed avionics system offers a completely decentralized solution where sub-tasks of a function are hosted by different hardware devices. An update or change of one service does not affect the operation of the entire system as long as it follows the same input and output relations. In the system design, a Docker container is adopted for the microservices architecture and the multi-layer platform is implemented following the edge-fog-cloud computing paradigm. The \textit{Lightweight IoT based Smart Public Safety (LISPS)} system architecture, which utilizes microservices-enabled private blockchain network to secure navigation services, provides secured data sharing. According to functionality and task completion, all containerized microservices are divided into four types and are deployed both at the edge and fog layer.

\subsection{Safety Application Services}
These microservices provide smart navigation application functions, such as data processing, object identification, movement features extraction, anomalous behavior recognition, and alert actions. Real-time avionics streams are generated by ADS-B messages and transmitted to edge microservices for feature extraction. Lower level features are transferred to fog nodes for data aggregation and higher level analytic services, such as route pattern recognition and anomalous event detection \cite{wei2007game}. The light-weight Hyper Text Transfer Protocol (HTTP) webservice is deployed at the host and is responsible for data transfer between the edge and fog. Given different capacities of the host platform, the smart navigation services consist of edge services and fog services.

\begin{figure}[t]
    \centering
        \includegraphics[width=0.48\textwidth]{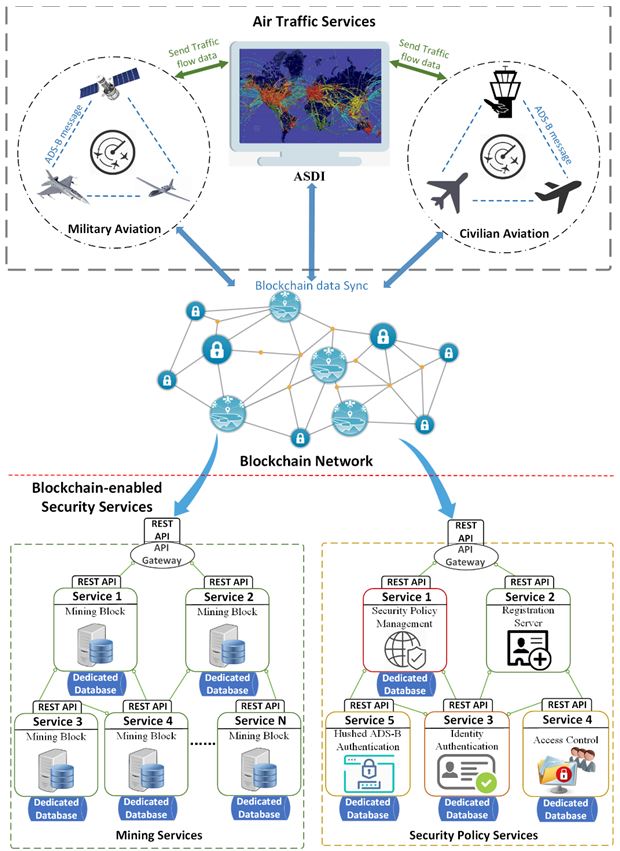}
    \caption{Systems Architecture of Blockchain ADS-B.}
    \label{fig:arch}
    \vspace{-10pt}
\end{figure}

\emph{Network Application Services on the Edge}: To reduce network communication latency and overload, raw ADS-B messages captured by navigation equipments are processed by edge devices that conduct low-level feature extraction tasks, such as aircraft identification and trajectories. The key function of the features extraction task can be decoupled into multiple microservices that are deployed on a single or multiple edge devices and work cooperatively, as shown by Fig. \ref{fig:arch}. In the aircraft tracking and identification (e.g., see \cite{blasch2004ten}) microservice, multiple frames are checked every second for feature extraction through a lightweight convolutional neural network (L-CNN) algorithm \cite{nikouei2018realb}. Then, a queue is maintained to track the aircraft within a airspace box, based on a lightweight, application. Finally, movement-based features, such as the speed and direction changes, are extracted for each aircraft in the data frame. The features for each aircraft of interest are put into a dictionary format where the key is the object first detection time and the value consists of all the features. The extracted lower level feature is converted to a JSON string and transferred to the fog layer using HTTP protocol communication channel.

\emph{Network Application Services on the Fog}: After extracted feature data have been merged by the fog platform through data streaming services, the higher-level feature contextualization and smart decision making for a surveillance system are available to support intelligent analytic functions. Prior to the decision making process, the data should be contextualized given significant factors, like the time of the day, the location of the signal, and the security level of the ADS-B system. After the contextualization, the features of each detected aircraft in the frame are divided and separately processed by data fusion logic microservices. Given pre-defined logic rules, the suspicious activity level regarding an aircraft route anomaly is returned. According to a threshold defined by the system air traffic controller (ATC) based on the past experiences, the decision making is performed using a suspicious activity level (e.g., 80\%). A text message is sent to the ATC if suspicion level of an individual aircraft is beyond the pre-set threshold.

\subsection{Blockchain-enabled Security Services}

In the LISPS system, security policy covers two main aspects: identity management and secured data accessing. An identity management mechanism ensures a new node enrollment process in the permissioned blockchain network. Only authorized participants could be recognized by entities of the network and perform blockchain services, such as mining blocks, sending transactions and deploying smart contracts. Compared to a public blockchain network, the \emph{permissioned blockchain network} achieves higher efficiency in consensus operation, and more secure by limiting participants and clearly defining security policies. The secured data accessing service acts as a fundamental service pool, which includes three main clusters: access control services, security policies services and mining services. All the entities on the permissioned blockchain network are implemented as containers, which perform blockchain services independently on the host devices. The containerized microservices could be categorized as miners or non-mining nodes given the computation power of the host devices.

Utilizing the microservices architecture, the security policy functions are decoupled into multiple microservices and deployed on distributed computing devices. These decentralized security microservices work as a service cluster to offer a scalable, flexible, and lightweight data sharing and access control mechanisms for the LISPS system. An entity registration process is performed by the registration microservices that associate entity’s unique blockchain account address with a Virtual ID (VID). The identity authentication microservices expose RESTful APIs to other microservices-enabled providers for referring identity verification results.

The security microservices act as data and security service managers who deploy the smart contracts that encapsulate identity authentication and the access control policies. After the smart contracts have been deployed successfully on the blockchain network, they become visible to the entire network. The authorized participants could interact with smart contract through the Remote Procedure Call (RPC) interfaces. The access control microservices encapsulate access control model and perform access right validation during service granting process on smart surveillance service providers.

\subsection{Trajectory Simulation Services}

The trajectory simulation services combine multiple individual simulations models and techniques to implementation/model development stages of the simulation. The simulation offers an enhanced representation of the system including: 

\begin{itemize}
    \item \emph{System dynamics simulation} is implemented as two microservices: sensor dynamics and system dynamics. The sensor dynamics simulation service could evaluate behavior of edge applications services and output estimates for sensor measurements and access control change. The system dynamics simulation services analyze the nonlinear behaviors of fog computing system to perform reconfiguration of the simulation models and system services. 
    \item \emph{Agent-based simulation} simulates the action and interaction of autonomous-given behavior recognition results from smart surveillance. Behavior recognition based on extracted lower features allows the assessment of individual aircraft in the ADS-B reporting, and those individual agents will be analyzed by agent-based simulation to generate estimate for discrete event simulation.
    \item \emph{Discrete-event simulation} combines estimates from agent-based model and a discrete sequence of events in time to model the operation of system. The event algorithm analyzes those multiple objects and assesses an event scenario during a certain period of time. In a discrete-event simulation, the event could refer to actions of aircraft in the airspace. The discrete-event simulation estimates coordinate agent-based services for improving accuracy.
\end{itemize}

Discrete event simulation and agent-based simulation microservices cooperate with each other under close feedback loop to assist management services and mining services.

\section{Blockchain Avionics Example}
\label{fuzzy}

\begin{figure}[t]
    \centering
        \includegraphics[width=0.48\textwidth]{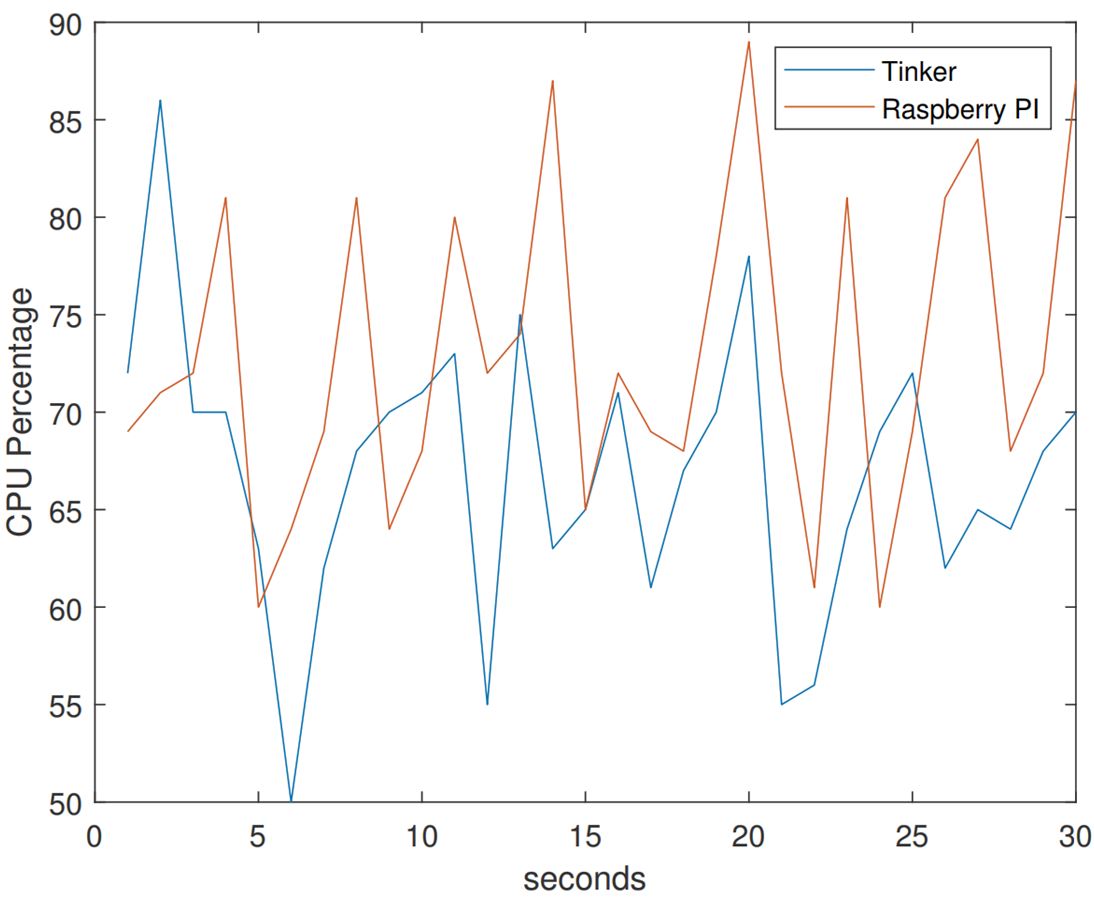}
    \caption{CPU usage percentage in the edge units.}
    \label{fig:cpu}
    \vspace{-10pt}
\end{figure}

A concept-of-proof prototype system has been implemented on a real physical network environment including a safety application and blockchain-enabled security services using ADS-B. The applications are decoupled as multiple microservices that are developed as Docker images, and are deployed both on fog and edge computing platform. A Raspberry PI acts as an edge device while a laptop is adopted as the fog node. Key metrics are assessed for performance and effectiveness over awareness \cite{salerno2005evaluating}, support \cite{blasch2011information}, veracity \cite{insaurralde2017veracity}, and relevance \cite{blasch2019uncertainty}.

Using the ADS-B, blockchain, and container-based systems; we augment the contextualization models of both the quantitative (processing needs) and the qualitative (social needs) for the system. Preliminary results demonstrate that the system performs in real time with edge units is utilized at 80\%. Figure \ref{fig:cpu} shows the advantage of the trajectory simulation as the dynamic prediction models, agent-based approach to container use, and simulations results direct the fog services which reduces the processing time for the ADS-B average analytics over a interval of $15 \times 100$ messages.  

\begin{figure}[t]
    \centering
        \includegraphics[width=0.48\textwidth]{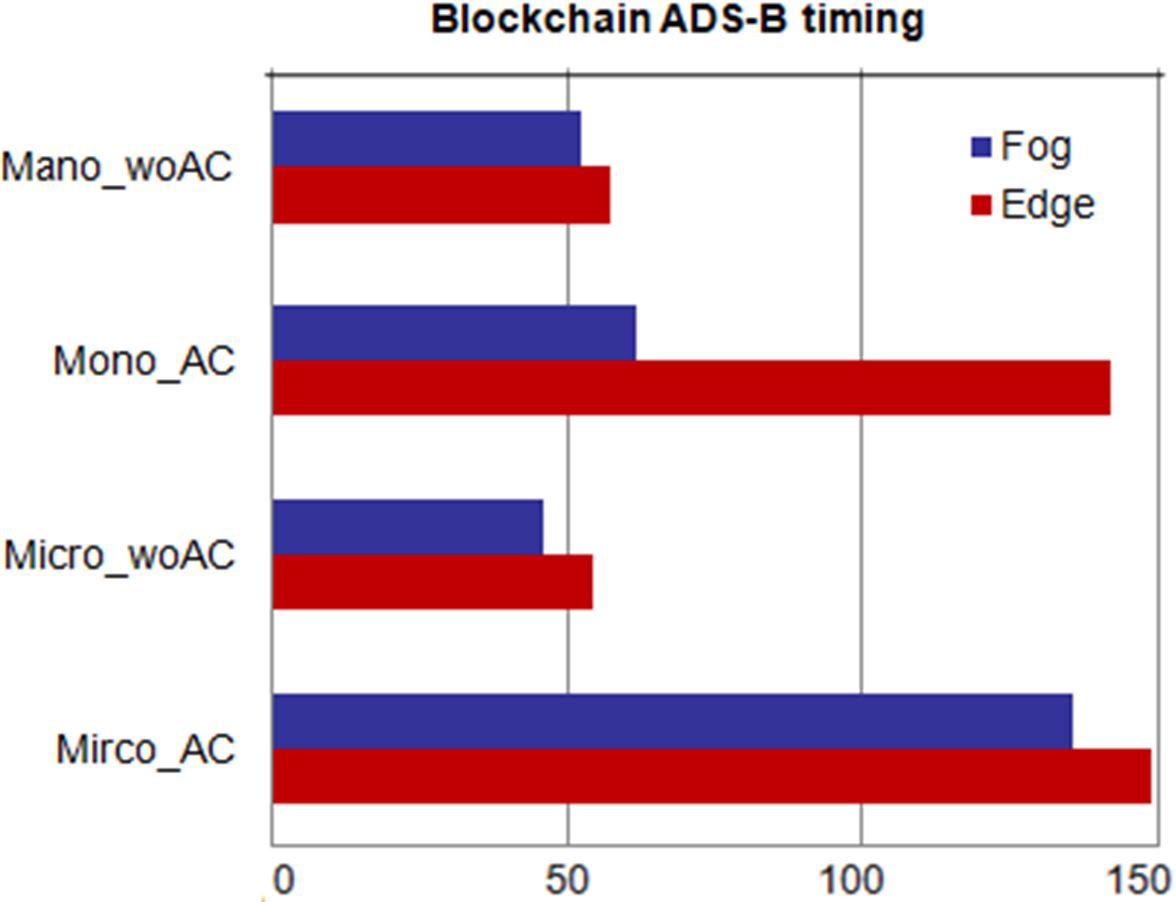}
    \caption{Network latency of BlendCAC.}
    \label{fig:net}
    \vspace{-10pt}
\end{figure}

The blockchain-enabled security services are built on a permissioned blockchain network including six miners and six non-miners participants. Two miners are deployed on a laptop and other four miners are distributed on four desktops. Each miner just use one CPU core to ensure they have same computation resource. Two desktop and four Raspberry PIs are non-miners to run security microservices. The security microservices are developed as Docker containers providing both Go-Ethereum as client application to interact with blockchain network and Flask-based webservice to handle security service request from user. To evaluate the performance of the microservices-based security mechanism, a service access experiment is carried out on a physical network environment which includes 3 Raspberry PIs and 2 desktops. One Raspberry PI works as a client to send service request, while server side is the navigation service provider, which has been both hosted on edge (Raspberry PI) and fog (desktop) nodes. A blockchain enabled capability based access control (BlendCAC) scheme [31, 90] is selected to enforce the access control policies.

To evaluate the overall network latency incurred by the microservices architecture, a comprehensive test has been performed on two service architectures: the microservices architecture (Micro App) and the monolithic framework (Mono App). Figure \ref{fig:net} shows the overall network latency incurred and compares the execution time of the BlendCAC and a benchmark without any access control enforcement on two service architecture. Considering the scenarios without access control enforcement, microservices and the monolithic framework have almost the same performance both on edge ($\sim 50$ ms) and fog platform ($\sim 48$ ms). However, when it comes to BlendCAC scenario, the experiments on two architectures show different communication latencies between the fog and edge platforms. Although microservices incurs more network latency, which is 75 ms on the fog side and 7 ms on edge side, it still brings benefits to the distributed ADS-B navigation-based systems, such as loosely coupled dependence, easy service deployment, and cross-platform interoperability.

\section{Conclusions}
\label{conc}

Future avionics systems using ADS-B require enhanced security to mitigate various attacks on the hardware, network, and software. This paper presents a blockchain with smart contract as a method to secure navigation communications, with little latency.  A microservice approach was presented using the advantages of edge and fog computing.

Future developments would include testing on a UAV transmitting and receiving ADS-B signals from both a cooperative and a non-cooperative aircraft.

\section*{ACKNOWLEDGEMENTS}

The views and conclusions contained herein are those of the authors and should not be interpreted as necessarily representing the official policies, either expressed or implied, of Air Force Research Laboratory, or the U.S. Government.

\ifCLASSOPTIONcaptionsoff
  \newpage
\fi

\bibliographystyle{IEEEtranS}
\bibliography{L-CNN}

\vfill

\end{document}